\documentclass[11pt]{article}
\usepackage{epsfig}
\usepackage{hyperref}

\begin{document}

\title{An arbitrary Lagrangian-Eulerian formulation for the numerical simulation of flow patterns generated by the hydromedusa \textit{Aequorea victoria}}

\author{Mehmet SAHIN  and Kamran MOHSENI\\
\\
Department of Aerospace Engineering Sciences,\\
University of Colorado, Boulder,\\
Colorado, 80309, USA}

\maketitle

\begin{abstract}
A new geometrically conservative arbitrary Lagrangian-Eulerian (ALE) formulation is presented for the moving boundary problems in the swirl-free cylindrical coordinates. The governing equations are multiplied with the radial distance and integrated over arbitrary moving Lagrangian-Eulerian quadrilateral elements. Therefore, the continuity and the geometric conservation equations take very simple form similar to those of the Cartesian coordinates.
The continuity equation is satisfied exactly within each element and a special attention is given to satisfy the geometric conservation law (GCL) at the discrete level. The equation of motion of a deforming body is solved in addition to the Navier-Stokes equations in a fully-coupled form. The mesh deformation is achieved by solving the linear elasticity equation at each time level while avoiding remeshing in order to enhance numerical robustness. The resulting algebraic linear systems are solved using an ILU(k) preconditioned GMRES method provided by the PETSc library. The present ALE method is validated for the steady and oscillatory flow around a sphere in a cylindrical tube and applied to the investigation of the flow patterns around a free-swimming hydromedusa \textit{Aequorea victoria} (crystal jellyfish). The calculations for the hydromedusa indicate the shed of the opposite signed vortex rings very close to each other and the formation of large induced velocities along the line of interaction while the ring vortices moving away from the hydromedusa.
In addition, the propulsion efficiency of the free-swimming hydromedusa is computed and its value is compared with values from the literature for several other species. The fluid dynamics video presented here shows the time variation of the instantaneous three-dimensional vorticity isosurfaces around a free-swimming hydromedusa \textit{Aequorea victoria}.
\end{abstract}

\section{Introduction}
The present dynamics video presented here shows the time variation of the instantaneous three-dimensional vorticity isosurfaces around a free-swimming hydromedusa \textit{Aequorea victoria}. To solve the flow pattern around the free-swimming hydromedusa \textit{Aequorea victoria}, a new geometrically conservative arbitrary Lagrangian-Eulerian (ALE) formulation presented in \cite{SaMo09} has been used.
The maximum bell radius of the medusa is approximately $2.3$ cm and the period of one cycle T is being approximately equal to 1.16 seconds. To compute the velocity of the medusa, the equation of motion is solved in addition to the Navier–Stokes equations in a fully coupled form. The calculations are carried out on high resolution computational meshes: a coarse mesh M1 with 63,099 vertices and
62,610 quadrilateral elements and a fine mesh M2 with 205,714 vertices and 204,784 quadrilateral elements. The computed average swimming velocity is computed to be $1.453$ cm/s and $1.462$ cm/s for the meshes M1 and M2, respectively. Based on the average medusa velocity on mesh M2 and the maximum
bell diameter, the dimensionless parameters Reynolds and Strouhal numbers are computed to be 672 and 8.47, respectively. The details of the present work can be found in the papers listed in the references.

\begin{figure}[h]
\begin{picture}(0,320)(-0,0)
\scalebox{0.1}{\includegraphics{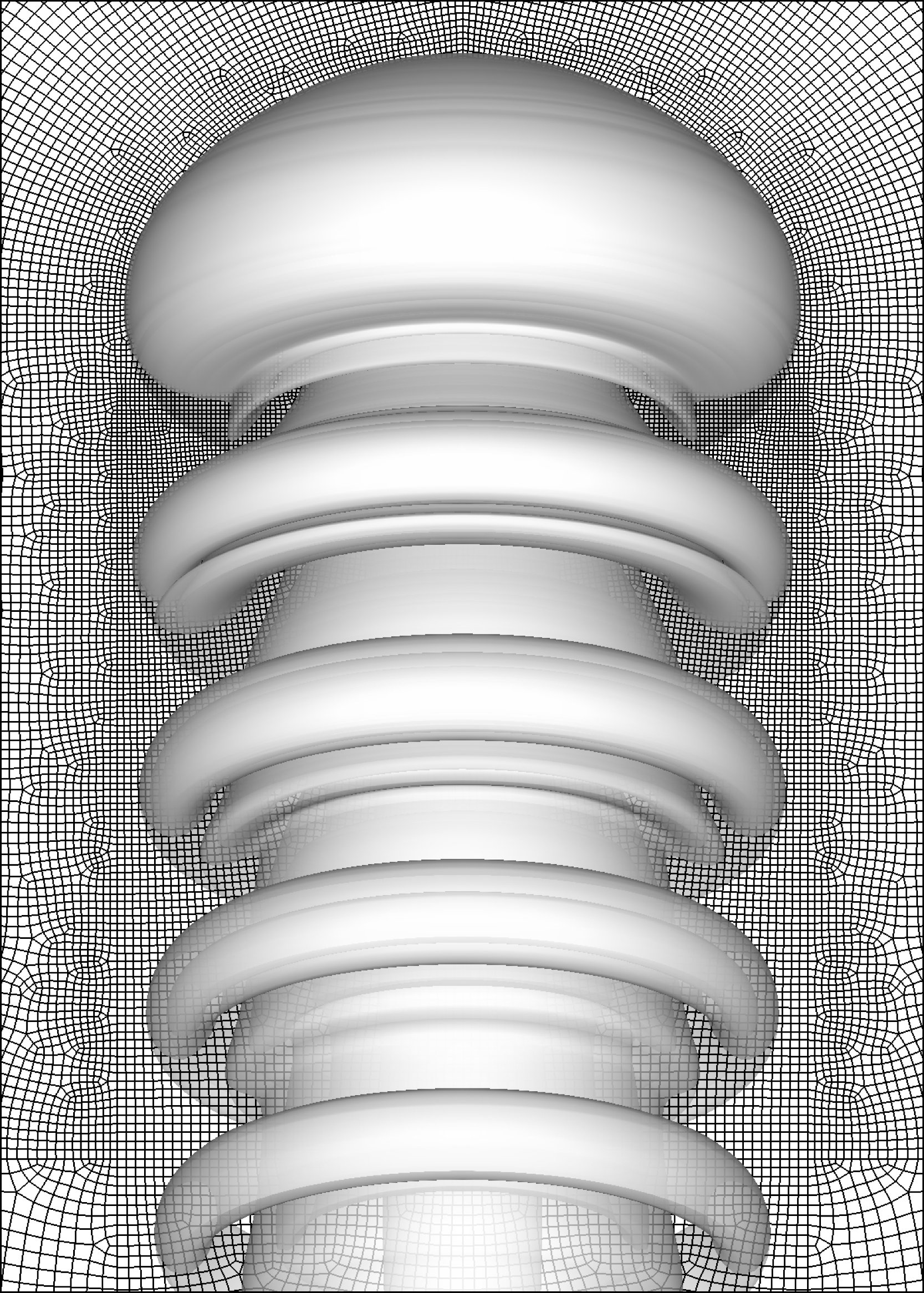}}[a]\hspace{6cm}\scalebox{0.1}{\includegraphics{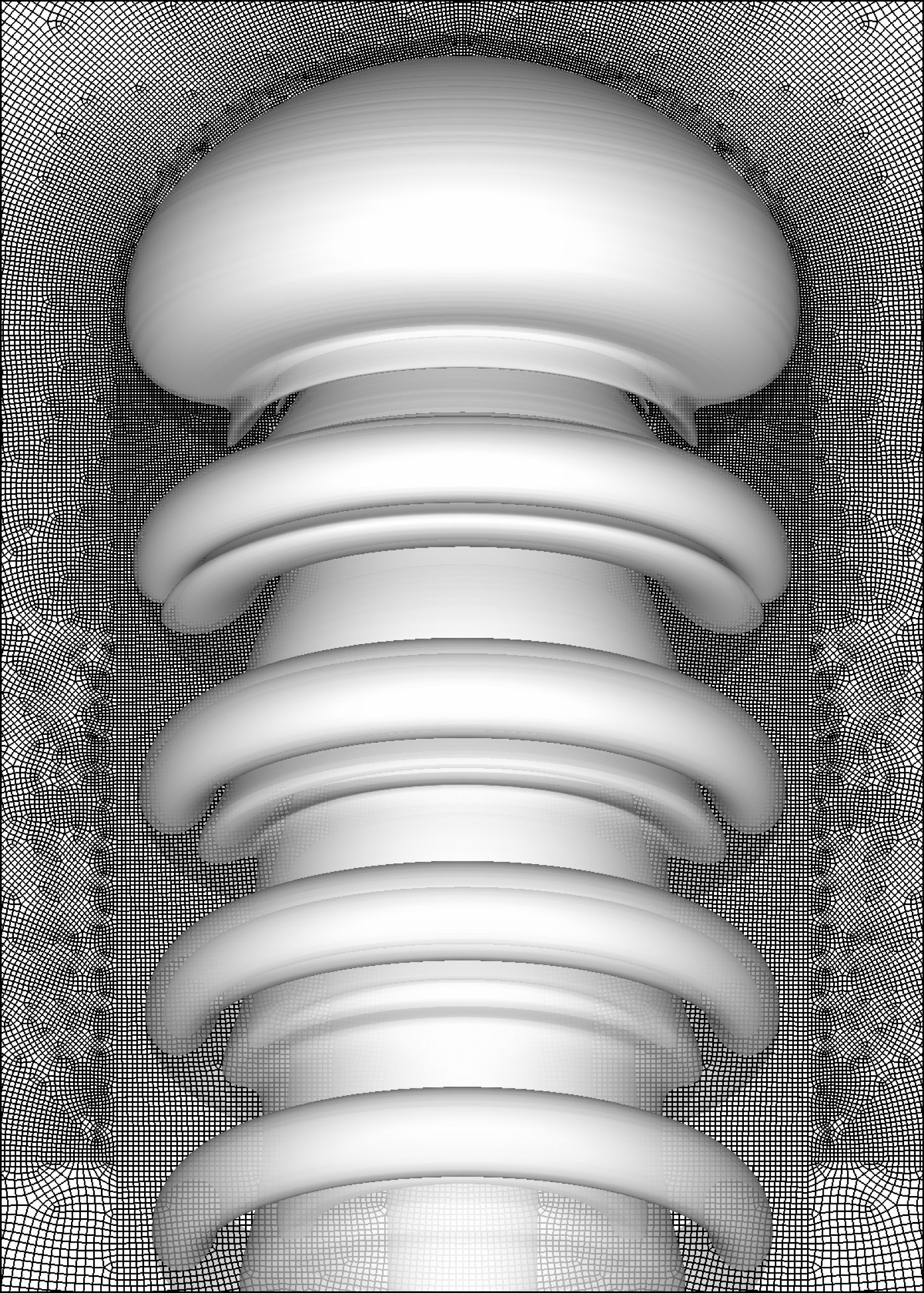}}[b]
\end{picture}
\caption{The mesh convergence is given for the wake structure behind a free-swimming hydromedusa \textit{Aequorea victoria} on meshes M1 [a] and M2 [b].} \label{FigWake}
\end{figure}


\begin{thebibliography}{87}
\bibitem{SaMo09}
M. Sahin and K. Mohseni,  An Arbitrary Lagrangian-Eulerian Formulation for the Numerical Simulation of Flow Patterns Generated by the Hydromedusa Aequorea Victoria. \textit{J. Comput. Phys.}, (2009), 228:4588-4605.

\bibitem{SaMoCo09}
M. Sahin, K. Mohseni and S. Colin, The Numerical Comparison of Flow Patterns and Propulsive Performances for the Hydromedusae Sarsia Tubulosa and Aequorea Victoria. \textit{J. Exp. Biol.}, (2009), 212:2656-2667.
\end{thebibliography}
\end{document}